\documentstyle[axodraw,aps,12pt]{revtex}
\begin{document}
\vskip0.5cm
\begin{center}
{\Large \bf  Charged Higgs Associated Production with $W$ at Linear Collider
 }
  \vskip1cm {\large Shou Hua Zhu 
}
  \vskip0.2cm
  CCAST (World Laboratory),
P. O.\ Box 8730, Beijing, 100080, P. R. China  \\
\vskip0.2cm
 Institute of Theoretical Physics, Academia Sinica, P.O. Box
 2735, Beijing 100080, P.R. China \\ 

  \vskip1.8cm
  {\large\bf ABSTRACT \\[10pt]} \parbox[t]{\textwidth}{
  We have calculated the cross section of the 
  process $e^+e^- \rightarrow W^{\pm}H^{\mp}$ at the linear
  collider in the supersymmetrical two-higgs-doublet model. 
  We find that the cross section decreases
  rapidly with the increment of $\tan\beta$, and which can
  reach 
  almost $1$ $fb$ 
  for the favorable parameters.
  }
  
  \vskip1cm

\end{center}

PACS number(s): 12.15.Lk,  12.60.Jv, 14.80.Cp

Keyword(s): Higgs Physics, Two-Higgs-doublet Model, Linear Collider 
\newpage

 
\section{Introduction} 
\setcounter{equation}{0} 
 
The Higgs boson is the missing piece and also the least known one of 
the standard model (SM) and beyond. 
Under the framework 
of the  minimal supersymmetrical
standard model (MSSM) \cite{MSSM}, there contain charged Higgs bosons, which
are one of the most distinguished signal compared with SM. To find this kind of
charged Higgs boson and study its properties are  
one of the primary goals of the present 
and next generation of colliders.

The production of the charged Higgs boson at hadron colliders can 
mainly via
top quark decay if kinetically allowed; recently, we have studied single
charged Higgs production through $bg$ channel and found this channel
could be helpful in finding the charged Higgs \cite{bg}. 
Otherwise, the heavy
charged Higgs would be very hard to produce at hadron colliders.
At linear collider, the primary mechanism
to produce charged Higgs bosons are $e^+e^- \rightarrow H^+H^-$
\cite{EEHH} or 
$\gamma\gamma \rightarrow H^+H^-$ \cite{AAHH}. 
However, the production rates for both
processes drop rapidly with the increment of the charged Higgs mass.
As a complementary channel in discovering charged Higgs,
although suffering from the low rate (this channel is the one loop process),
$e^+e^- \rightarrow W^{\pm}H^{\mp}$ channel 
is important in studying the $\gamma-W^{\pm}-
H^{\mp}$ and $Z-W^{\pm}-
H^{\mp}$ vertexes.
In fact, these two vertexes have been studied through the decay processes
of the charged Higgs boson \cite{Vertex}. 

In literatures, there are many works on neutral Higgs associated 
production with
gauge bosons ($\gamma$, $Z$ and $W$) 
both on hadron and linear colliders \cite{NEUTRAL}, which are potentially
useful in dealing with the backgrounds. 
However, the charged Higgs
bosons associated production with $W$ at linear collider
is deserved a study, which will be the topic of this paper.
At linear collider, there is an alternative collision mode:
$\gamma\gamma$ collision, which is commonly thought the
most suitable place to study the Higgs properties. The $W^\pm
H^\mp$ pair production in this collision mode will be presented in the
forthcoming work \cite{Zhu}.

\section{FORMALISM OF Charged HIGGS BOSON PRODUCTION ASSOCIATED WITH $W$ }

The Feynman diagrams of the  process
$e^{+}(p_1) e^- (p_2)\rightarrow H^{\pm} (k_1) W^{\mp} (k_2)$ 
are shown in Fig 1, and the 
corresponding Feynman rules could be found in Ref.
\cite{Hunter}. We perform 
the calculation in the 't Hooft-Feynman gauge and use dimensional 
regularization to all the ultraviolet divergence in the virtual 
loop contributions. 
 
As usual, we define the mandelstam variables as 
\begin{eqnarray} 
S&=&(p_1+p_2)^2=(k_1+k_2)^2\nonumber \\ 
T&=&(p_1-k_1)^2=(p_2-k_2)^2 \nonumber \\ 
U&=&(p_1-k_2)^2=(p_2-k_1)^2. 
\end{eqnarray} 
 
The amplitude of the process can be written as
\begin{eqnarray}
M&=& \bar U(p_2) \left(
f_1 \rlap/\epsilon P_R +
f_2 \rlap/\epsilon P_L +
\rlap/k_2 \left(
(f_3 P_R 
+f_4 P_L) p_1 . \epsilon+
(f_5 P_R 
+f_6 P_L) p_2 . \epsilon \right)
\right) V(p_1),
\end{eqnarray}
where $\epsilon$ is the polarization vector of W boson and
$P_{L,R}=(1\mp \gamma_5)/2$.
The form factors $f_i$ could be expressed as
\begin{eqnarray}
f_i=\sum_{j=1}^{11} f_i^{(j)},
\label{EQ1}
\end{eqnarray}
where $j$ corresponds to the contributions from the diagrams index
in Fig. 1. For simplicity, if possible, the form factors
$f_i^{(j)}$ are divided into 
$f_i^{(j),\gamma}$ and $f_i^{(j),Z}$, which are from
the s-channel $\gamma$ and $Z$ contributions, respectively;
for diagram (2) in Fig. 1, the form factors are further divided
into fermionic and bosonic contributions. The non-vanishing form factors are 
explicitly presented in Appendix.

The cross section of  process 
$e^{+}(p_1) e^{-}(p_2)\rightarrow H^{\pm} (k_1) W^{\mp} (k_2)$ 
 can be written as 
\begin{eqnarray} 
\sigma =\int_{T_{min}}^{T_{max} }{1\over 16\pi S^2}\overline 
{\sum_{ spins} }\left|  M\right|^2 d T 
\end{eqnarray} 
with 
\begin{eqnarray} 
T_{min}&=&{1\over 2} (m_{H^\pm}^2+m_W^2-S-\sqrt{ 
   (S-(m_{H^\pm}+m_W)^2)(S-(m_{H^\pm}-m_W)^2)}) \nonumber \\ 
T_{max}&=&{1\over 2} (m_{H^\pm}^2+m_W^2-S+\sqrt{ 
   (S-(m_{H^\pm}+m_W)^2)(S-(m_{H^\pm}-m_W)^2)})  
\end{eqnarray} 
where the bar over the summation recalls average over initial fermions spins. 

\section{NUMERICAL RESULTS AND DISCUSSIONS}

In the following we present some numerical 
results of  charged
Higgs boson associated 
production with $W$ in the process of
$ e^+e^- \rightarrow H^\pm W^\mp$. In our calculations,
we choose $m_W=80.33 GeV$, $m_Z=91.187 GeV$,
$m_t=176 GeV$, $m_b=4.5 GeV$ and $\alpha=1/128$. 
For simplicity, we choose the parameters
of the Higgs sectors which satisfied the MSSM relations, i.e., there are two
independent parameters $\tan\beta$ and $m_{H^\pm}$ \cite{Hunter}.

In Fig. 2, the $W^{\pm} H^{\mp} $ pair production cross sections as a function
of the $m_{H^{\pm}}$ at linear collider are presented, 
where $\sqrt{S}=500 GeV$. From the figures, we can see that the cross sections
drops rapidly with the increment of $\tan\beta$, and for $\tan\beta=2$ and
$m_{H^\pm}=100 \sim 250$ GeV,
the cross sections are insensitive to the mass of the charged Higgs boson,
which is around 0.2 fb. We note that the peak around 180 GeV is due to the
threshold effect that the decay channel of the charged Higgs boson to
top and bottom are opened.

In Fig. 3, the $W^{\pm} H^{\mp} $ pair production cross sections as a function
of the $\sqrt{S}$ are presented for $m_{H^\pm}=200$ GeV. We can
see that there is a peak around $400$ GeV, which is the consequence of the
competitive between the enlargement of phase space and the s-channel
suppression. For three cases of $\tan\beta$, the cross sections of
small $\tan\beta$ are larger than that of large $\tan\beta$, which can
reach 1 fb for favorable parameters. 

To summarize, we have calculated the cross section of the 
process $e^+e^- \rightarrow W^{\pm}H^{\mp}$ at the next linear
collider in the supersymmetrical two-higgs-doublet model. 
We find that the cross section decreases
rapidly with the increment of $\tan\beta$, and which can
reach almost $1$ $fb$ 
for the favorable parameters.
We note that in supersymmetrical framework including MSSM, 
the s-particles can also contribute
to this process, which will be studied in the near future.

\section*{Acknowledgments}
This work was supported in part by the post doctoral foundation
of China and the author gratefully acknowledges the
support of  K.C. Wong Education Foundation, Hong Kong.


\newpage 
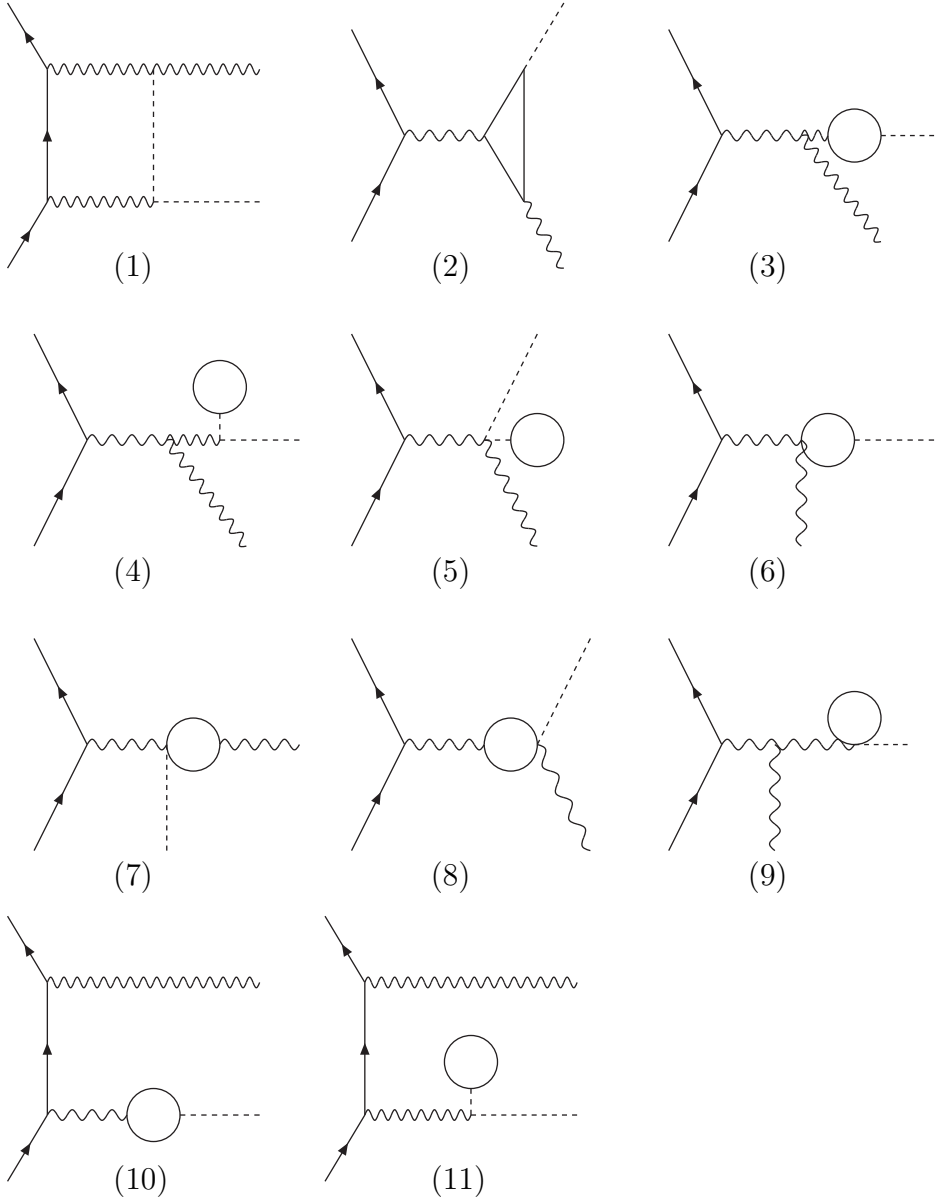
\begin{figure}
\begin{picture}(400,100)
\Text(40,0)[l]{(1)}
\Text(160,0)[l]{(2)}
\Text(280,0)[l]{(3)}
\SetOffset(0,0)
\ArrowLine(0,0)(15,25)
\ArrowLine(15,25)(15,75)
\ArrowLine(15,75)(0,100)

\Photon(15,75)(95,75){2}{16}
\Photon(15,25)(55,25){2}{8}
\DashLine(55,25)(55,75){2}
\DashLine(55,25)(95,25){2}

\SetOffset(120,0)
\ArrowLine(10,10)(30,50)
\ArrowLine(30,50)(10,90)

\Photon(30,50)(60,50){2}{4}

\Line(60,50)(75,75)
\Line(60,50)(75,25)
\Line(75,75)(75,25)

\DashLine(75,75)(90,100){2}
\Photon(75,25)(90,0){2}{4}

\SetOffset(240,0)
\ArrowLine(10,10)(30,50)
\ArrowLine(30,50)(10,90)

\Photon(30,50)(60,50){2}{4}
\Photon(60,50)(90,10){2}{8}
\Photon(60,50)(70,50){2}{2}
\CArc(80,50)(10,0,180)
\CArc(80,50)(10,180,360)
\DashLine(90,50)(110,50){2}
\end{picture}

\vskip 0.5cm
\begin{picture}(400,100)
\Text(40,0)[l]{(4)}
\Text(160,0)[l]{(5)}
\Text(280,0)[l]{(6)}
\SetOffset(0,0)
\ArrowLine(10,10)(30,50)
\ArrowLine(30,50)(10,90)

\Photon(30,50)(60,50){2}{4}
\Photon(60,50)(90,10){2}{8}
\Photon(60,50)(80,50){2}{4}
\DashLine(80,50)(80,60){2}
\CArc(80,70)(10,0,180)
\CArc(80,70)(10,180,360)
\DashLine(80,50)(110,50){2}

\SetOffset(120,0)
\ArrowLine(10,10)(30,50)
\ArrowLine(30,50)(10,90)
\Photon(30,50)(60,50){2}{4}
\DashLine(60,50)(80,90){2}
\Photon(60,50)(80,10){2}{6}
\DashLine(60,50)(70,50){2}
\CArc(80,50)(10,0,180)
\CArc(80,50)(10,180,360)

\SetOffset(240,0)
\ArrowLine(10,10)(30,50)
\ArrowLine(30,50)(10,90)
\Photon(30,50)(60,50){2}{4}
\CArc(70,50)(10,0,180)
\CArc(70,50)(10,180,360)
\DashLine(80,50)(110,50){2}
\Photon(60,50)(60,10){2}{4}
\end{picture}

\vskip 0.5cm
\begin{picture}(400,100)
\Text(40,0)[l]{(7)}
\Text(160,0)[l]{(8)}
\Text(280,0)[l]{(9)}
\SetOffset(0,0)
\ArrowLine(10,10)(30,50)
\ArrowLine(30,50)(10,90)
\Photon(30,50)(60,50){2}{4}
\CArc(70,50)(10,0,180)
\CArc(70,50)(10,180,360)
\DashLine(60,50)(60,10){2}
\Photon(80,50)(110,50){2}{4}

\SetOffset(120,0)
\ArrowLine(10,10)(30,50)
\ArrowLine(30,50)(10,90)
\Photon(30,50)(60,50){2}{4}
\CArc(70,50)(10,0,180)
\CArc(70,50)(10,180,360)
\DashLine(80,50)(100,90){2}
\Photon(80,50)(100,10){2}{4}

\SetOffset(240,0)
\ArrowLine(10,10)(30,50)
\ArrowLine(30,50)(10,90)
\Photon(30,50)(80,50){2}{6}
\CArc(80,60)(10,0,180)
\CArc(80,60)(10,180,360)
\DashLine(80,50)(100,50){2}
\Photon(50,50)(50,10){2}{4}
\end{picture}

\vskip 0.5cm
\begin{picture}(400,100)
\Text(40,0)[l]{(10)}
\Text(160,0)[l]{(11)}
\SetOffset(0,0)
\ArrowLine(0,0)(15,25)
\ArrowLine(15,25)(15,75)
\ArrowLine(15,75)(0,100)

\Photon(15,75)(95,75){2}{16}
\Photon(15,25)(45,25){2}{4}
\CArc(55,25)(10,0,180)
\CArc(55,25)(10,180,360)
\DashLine(65,25)(95,25){2}

\SetOffset(120,0)
\ArrowLine(0,0)(15,25)
\ArrowLine(15,25)(15,75)
\ArrowLine(15,75)(0,100)

\Photon(15,75)(95,75){2}{16}
\Photon(15,25)(55,25){2}{8}
\DashLine(55,25)(55,35){2}
\CArc(55,45)(10,0,180)
\CArc(55,45)(10,180,360)
\DashLine(55,25)(95,25){2}
\end{picture}
\caption[]{ The typical Feynman diagrams for the process
of $e^+e^- \rightarrow H^\pm W^\mp$.} 
\label{fig1}
\end{figure}

\begin{figure}
\epsfxsize= 18cm
\centerline{\epsffile{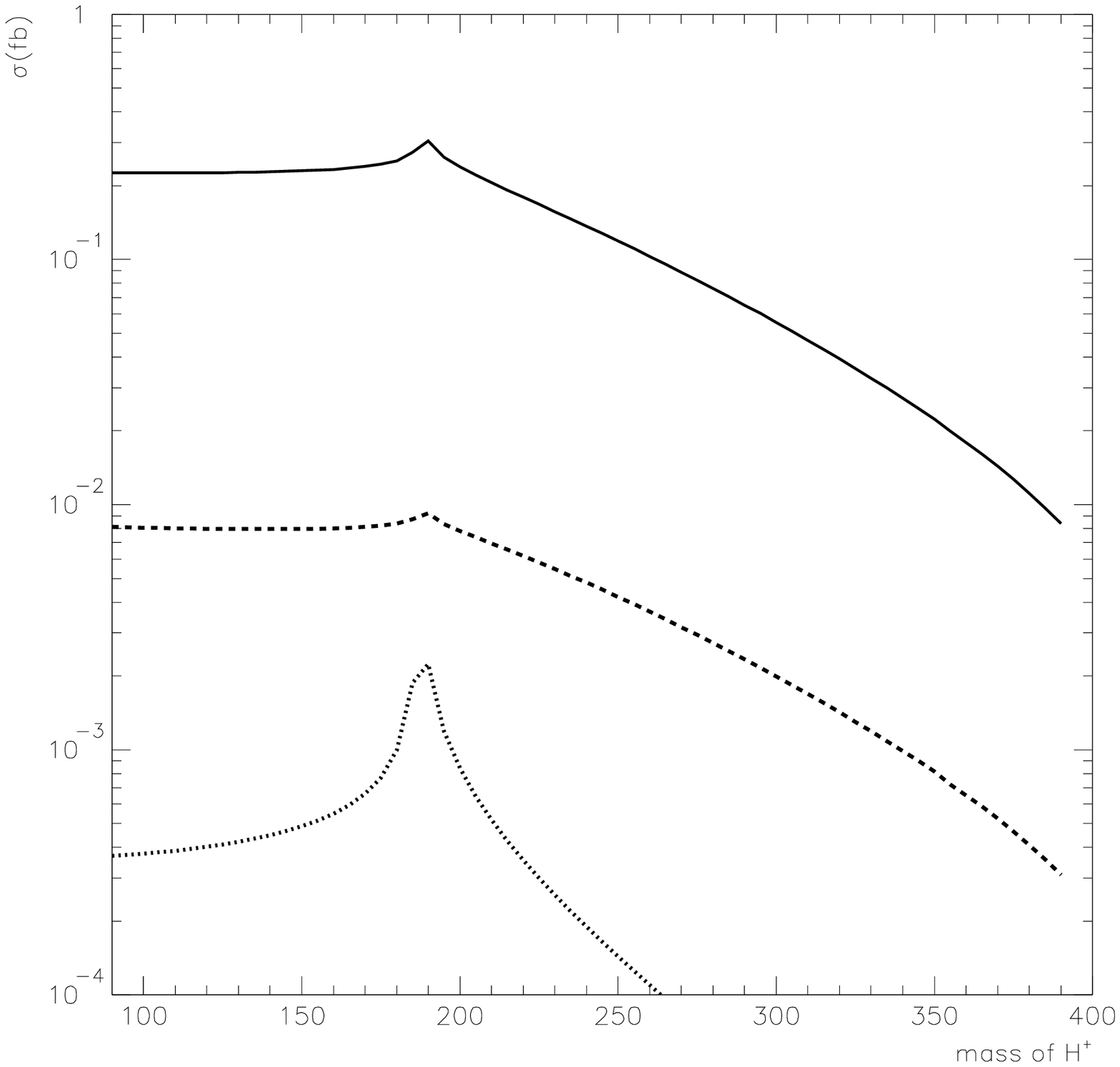}}
\caption[]{ 
The $W^{\pm} H^{\mp} $ pair production cross sections as a function
of the $m_{H^{\pm}}$ at linear collider, where $\sqrt{S}=500 GeV$, and
the solid, dashed and dotted lines represent $\tan\beta=2$, $10$ and
$40$, respectively.}
\label{fig2}
\end{figure}

\begin{figure}
\epsfxsize= 18cm
\centerline{\epsffile{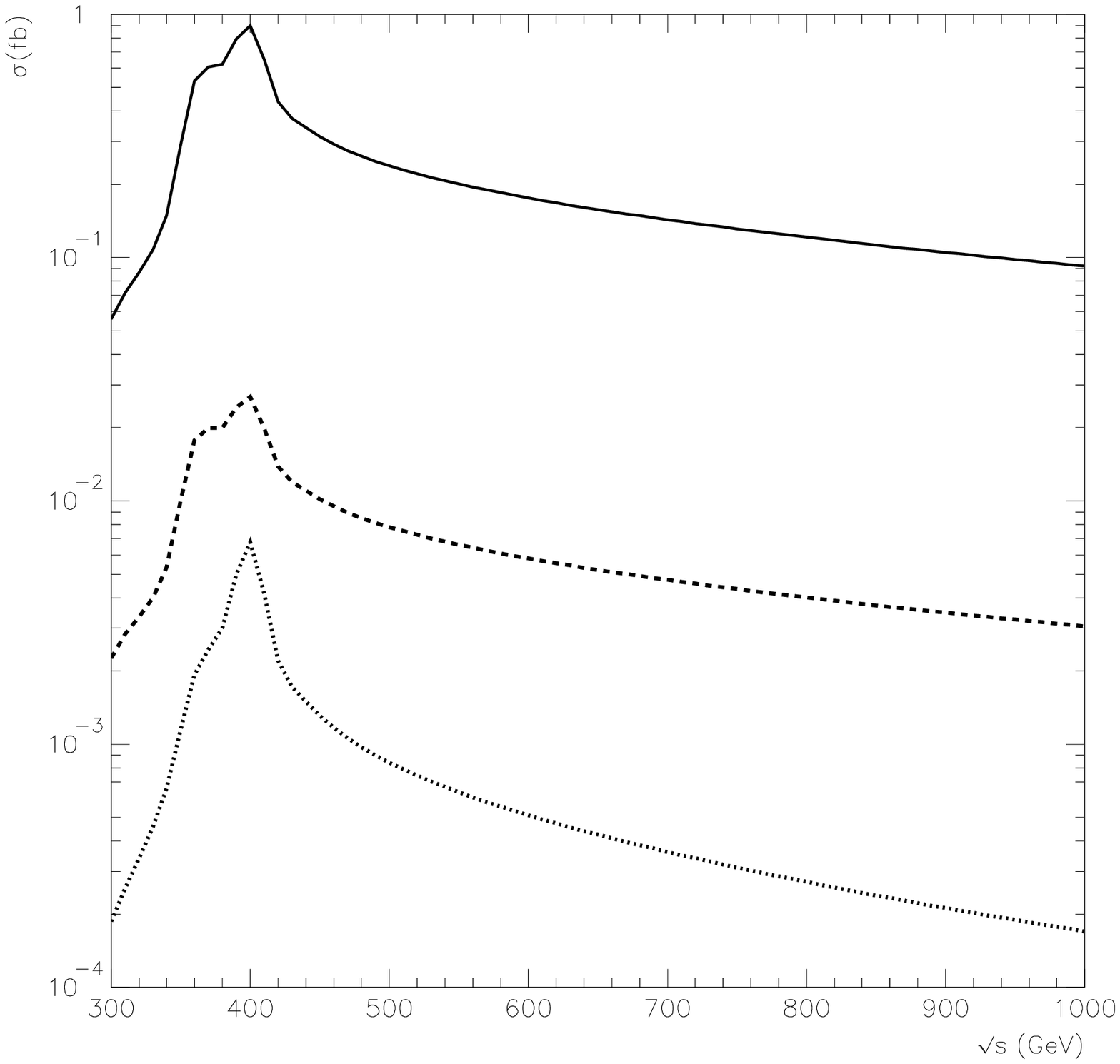}}
\caption[]{ 
The $W^{\pm} H^{\mp}$ pair production cross sections as a function
of the $\sqrt{S}$ at linear collider, where $m_{H^{\pm}}=200 GeV$, and
the solid, dashed and dotted lines represent $\tan\beta=2$, $10$ and
$40$, respectively.}
\label{fig3}
\end{figure}

\newpage

\section*{Appendix}
In this Appendix, the non-zero form factors defined in
Eq. \ref{EQ1} are explicitly written as
\begin{eqnarray}
f_{2}^{(1)}&=&
{\alpha^2 m_W \over 4 \sin^4\theta_W}
\cos(\alpha-\beta)
\sin(\alpha-\beta)
\left[
(T-m_H^2) D_0(1)
-(T-m_{h_0}^2) D_0(2)
\right.
\nonumber \\
&&\left.
+(T-m_{H^\pm}^2) \left( D_1(2)-D_1(1) \right)
+(T-m_{W}^2) \left( D_3(1)-D_3(2) \right)
\right],
\end{eqnarray}
\begin{eqnarray}
f_{6}^{(1)}&=&
{\alpha^2 m_W \over  \sin^4\theta_W}
\cos(\alpha-\beta)
\sin(\alpha-\beta)
\left(
D_3(1)-D_3(2)
\right),
\end{eqnarray}
\begin{eqnarray}
f_{1}^{(2),\gamma,F}&=&
{\alpha^2 \over m_W \sin^2\theta_W S}
\left[
-m_t^2 \cot\beta B_0(S,m_b^2,m_b^2)
+2 m_b^2 \tan\beta B_0(S,m_t^2,m_t^2) \right.
\nonumber \\
&&+m_t^2 (m_b^2-m_t^2-T) \cot\beta 
C_0(1)+
2 m_b^2 (m_b^2-m_t^2+U)\tan\beta 
C_0(2)
\nonumber \\
&&
+2 \left[
(m_t^2 \cot\beta (U-m_{H^\pm}^2)+ m_b^2 \tan\beta (U+m_{H^\pm}^2) \right]
C_1(2)
\nonumber \\
&&
- \left[
(m_t^2 \cot\beta (T+m_{H^\pm}^2)+ m_b^2 \tan\beta (T-m_{H^\pm}^2) \right]
C_1(1)
\nonumber \\
&&
-2 \left[
(m_t^2 \cot\beta (T-m_{W}^2)- m_b^2 \tan\beta (U+m_{W}^2) \right]
C_2(2)
\nonumber \\
&&
- \left[
(m_t^2 \cot\beta (T+m_{W}^2)+ m_b^2 \tan\beta (m_{W}^2-U) \right]
C_2(1)
\nonumber \\
&&
\left.
-2 
(m_t^2 \cot\beta + m_b^2 \tan\beta ) 
(2 C_{00}(2)-C_{00}(1))
\right],
\end{eqnarray}
\begin{eqnarray}
f_{2}^{(2),\gamma,F}&=&
f_{1}^{(2),\gamma,F}(T \leftrightarrow U),
\end{eqnarray}
\begin{eqnarray}
f_{3}^{(2),\gamma,F}&=&
{2 \alpha^2 \over m_W \sin^2\theta_W S}
\left[ 
2 m_b^2 \tan\beta C_0(2) 
\right. \nonumber \\
&&
+2 (2 m_b^2 \tan\beta +m_t^2 \cot\beta) C_1(2)
- m_t^2 \cot\beta C_1(1)
\nonumber \\
&&
+ m_b^2 \tan\beta (2 C_2(2) +C_2(1))
+(m_b^2 \tan\beta +m_t^2 \cot\beta) (2 C_{11}(2)-C_{11}(1))
\nonumber \\
&&
\left.
+(m_b^2 \tan\beta +m_t^2 \cot\beta) (2 C_{12}(2)-C_{12}(1))
\right],
\end{eqnarray}
\begin{eqnarray}
f_{4}^{(2),\gamma,F}&=&
{2 \alpha^2 \over m_W \sin^2\theta_W S}
\left[ 
- m_t^2 \cot\beta C_0(1) 
\right. \nonumber \\
&&
- (2 m_t^2 \cot\beta +m_b^2 \tan\beta) C_1(1)
+2 m_b^2 \tan\beta C_1(2)
\nonumber \\
&&
- m_t^2 \cot\beta (2 C_2(2) +C_2(1))
+(m_b^2 \tan\beta +m_t^2 \cot\beta) (2 C_{11}(2)-C_{11}(1))
\nonumber \\
&&
\left.
+(m_b^2 \tan\beta +m_t^2 \cot\beta) (2 C_{12}(2)-C_{12}(1))
\right],
\end{eqnarray}
\begin{eqnarray}
f_{5}^{(2),\gamma,F}&=&
f_{4}^{(2),\gamma,F},
\end{eqnarray}
\begin{eqnarray}
f_{6}^{(2),\gamma,F}&=&
f_{3}^{(2),\gamma,F},
\end{eqnarray}
\begin{eqnarray}
f_{1}^{(2),Z,F}&=&
{\alpha^2 \over 2 \sin^2\theta_W \cos^2\theta_W m_W (s-m_Z^2)}
\left[
\left( (3-2 \sin^2\theta_W) m_t^2 \cot\beta
+3 m_b^2 \tan\beta \right) B_0(S,m_b^2,m_b^2) \right.
\nonumber \\
&&
- 
\left( (3-4 \sin^2\theta_W) m_b^2 \tan\beta + 3 m_t^2 \cot\beta 
\right) B_0(S,m_t^2,m_t^2)
\nonumber \\
&&
+m_t^2 \left(
(T+m_t^2) \cot\beta (3-2 \sin^2\theta_W)+
2 m_b^2 \sin^2\theta_W \cot\beta
+3 m_b^2 \tan\beta \right) C_0(2)
\nonumber \\
&&
-m_b^2 \left(
(U+m_b^2) \tan\beta (3-4 \sin^2\theta_W)+
4 m_t^2 \sin^2\theta_W \tan\beta
+3 m_t^2 \cot\beta \right) C_0(1)
\nonumber \\
&&
+\left(
\cot\beta m_t^2 \left(-4 m_{H^\pm}^2 \sin^2\theta_W
-(3- 4 \sin^2\theta_W) U \right) 
- \tan\beta m_b^2 
(3-4 \sin^2\theta_W)
(m_{H^\pm}^2+U)
\right) C_1(2)
\nonumber \\
&&
-\left(
\tan\beta m_b^2 \left(-2 m_{H^\pm}^2 \sin^2\theta_W
+(3- 2 \sin^2\theta_W) T \right) 
- \cot\beta m_t^2 
(3-2 \sin^2\theta_W)
(m_{H^\pm}^2+T)
\right) C_1(1)
\nonumber \\
&&+\left(
\cot\beta m_t^2 \left(-4 T \sin^2\theta_W
-(3- 4 \sin^2\theta_W) m_W^2 \right) 
- \tan\beta m_b^2 
(3-4 \sin^2\theta_W)
(m_{W}^2+U)
\right) C_2(2)
\nonumber \\
&&
+\left(
\tan\beta m_b^2 \left(2 U \sin^2\theta_W
+(3- 2 \sin^2\theta_W) m_W^2 \right) 
+ \cot\beta m_t^2 
(3-2 \sin^2\theta_W)
(m_{W}^2+T)
\right) C_2(1)
\nonumber \\
&&\left.
+2 (m_t^2 \cot\beta+ m_b^2 \tan\beta)
\left(
(3-4 \sin^2\theta_W)C_{00}(2)-
(3-2 \sin^2\theta_W)C_{00}(1) \right)
\right],
\end{eqnarray}
\begin{eqnarray}
f_{2}^{(2),Z,F}&=&
{
-1+2 \sin^2\theta_W 
\over 
2 \sin^2\theta_W 
} 
f_{1}^{(2),Z,F}(T \leftrightarrow U),
\end{eqnarray}
\begin{eqnarray}
f_{3}^{(2),Z,F}&=&
{\alpha^2 \over m_W \sin^2\theta_W \cos^2\theta_W (S-m_Z^2)}
\left[ 
 m_b^2 \tan\beta (-3+4 \sin^2\theta_W) C_0(2) 
\right. \nonumber \\
&&
+ (2 m_b^2 \tan\beta +m_t^2 \cot\beta) (-3+4 \sin^2\theta_W) C_1(2)
- m_t^2 \cot\beta(-3+2 \sin^2\theta_W)  C_1(1)
\nonumber \\
&&
+ m_b^2 \tan\beta ((-3+4 \sin^2\theta_W) C_2(2) +2 C_2(1))
+(m_b^2 \tan\beta +m_t^2 \cot\beta)\times \nonumber \\
&&\left( (-3+4 \sin^2\theta_W) C_{11}(2)-
(-3+2 \sin^2\theta_W)C_{11}(1) \right)
\nonumber \\
&&
\left.
+(m_b^2 \tan\beta +m_t^2 \cot\beta) ((-3+4 \sin^2\theta_W) C_{12}(2)-
(-3+2 \sin^2\theta_W)C_{12}(1))
\right],
\end{eqnarray}
\begin{eqnarray}
f_{4}^{(2),Z,F}&=&
{ \alpha^2 (-1+2 \sin^2\theta_W) 
\over 2 m_W \cos^2\theta_W \sin^2\theta_W S}
\left[ 
 m_t^2 \cot\beta (3-2 \sin^2\theta_W)
  C_0(1) 
\right. \nonumber \\
&&
+(3-2 \sin^2\theta_W) (2 m_t^2 \cot\beta +m_b^2 \tan\beta) C_1(1)
+ m_b^2 \tan\beta (-3 + 4 \sin^2\theta_W) C_1(2)
\nonumber \\
&&- m_t^2 \cot\beta (4 C_2(2) +(-3+2 \sin^2\theta_W)C_2(1))
\nonumber \\
&&
+(m_b^2 \tan\beta +m_t^2 \cot\beta) ((-3+4 \sin^2\theta_W) C_{11}(2)
-(-3+2 \sin^2\theta_W) C_{11}(1))
\nonumber \\
&&
\left.
+(m_b^2 \tan\beta +m_t^2 \cot\beta) ((-3+4 \sin^2\theta_W) C_{12}(2)
-(-3+2 \sin^2\theta_W) C_{12}(1))
\right],
\end{eqnarray}
\begin{eqnarray}
f_{5}^{(2),Z,F}&=&
{2 \sin^2\theta_W \over 
-1+2 \sin^2\theta_W } f_{4}^{(2),Z,F},
\end{eqnarray}
\begin{eqnarray}
f_{6}^{(2),Z,F}&=&
{
-1+2 \sin^2\theta_W 
\over 
2 \sin^2\theta_W 
} f_{3}^{(2),Z,F},
\end{eqnarray}
\begin{eqnarray}
f_{1}^{(2),\gamma,B}&=&
{
\alpha^2 m_W \over 2 S \sin^2\theta_W }
\left[
-\cos(\alpha-\beta)
\left( 
(S+m_H^2) \sin(\alpha-\beta)+
m_Z^2 \cos(\alpha+\beta) \sin(2\beta)
\right) C_0(3)
\right.
\nonumber \\
&&
\sin(\alpha-\beta)
\left( 
(S+m_{h_0}^2) \cos(\alpha-\beta)+
m_Z^2 \sin(\alpha+\beta) \sin(2\beta)
\right) C_0(4)
\nonumber \\
&&
+ \cos(\alpha-\beta)\sin (\alpha-\beta) \left(
(2 m_{H^\pm}^2-T-U) \left(C_1(3)-C_1(4) \right)
\right.
\nonumber \\
&&
\left.
- (2 m_{W}^2-T-U) \left(C_2(3)-C_2(4) \right)
\right)
\nonumber \\
&&
+2 \sin (\alpha-\beta) \left(
-2 \cos (\alpha-\beta)
+\sec^2\theta_W \cos (\alpha+\beta) \cos (2\beta)\right)
C_{00}(5)
\nonumber \\
&&
+2 \cos (\alpha-\beta) \left(
2 \sin (\alpha-\beta)
+\sec^2\theta_W \cos (\alpha+\beta)\sin(2\beta) \right)
C_{00}(3)
\nonumber \\
&&
+2 \cos (\alpha-\beta) \left(
2 \sin (\alpha-\beta)
-\sec^2\theta_W \sin (\alpha+\beta) \cos (2\beta)\right)
C_{00}(6)
\nonumber \\
&&
\left.
+2 \sin (\alpha-\beta) \left(
-2 \cos (\alpha-\beta)
+\sec^2\theta_W \sin (\alpha+\beta)\sin(2\beta) \right)
C_{00}(4)
\right],
\end{eqnarray}
\begin{eqnarray}
f_{2}^{(2),\gamma,B}&=&
f_{1}^{(2),\gamma,B},
\end{eqnarray}
\begin{eqnarray}
f_{3}^{(2),\gamma,B}&=&
{
\alpha^2 m_W \over  S \sin^2\theta_W }
\left[
\sin(\alpha-\beta) \left(
2 \cos (\alpha-\beta)- \sec^2\theta_W \cos(2\beta)
\cos (\alpha+\beta) \right) \left(C_1(5)+C_{11}(5)
+C_{12}(5)\right)
\right.
\nonumber \\
&&
-\cos (\alpha-\beta) \cos (\alpha+\beta) \sin(2\beta)
\sec^2\theta_W C_1(3)
\nonumber \\
&&
+\cos(\alpha-\beta) \left(
-2 \sin (\alpha-\beta)+ \sec^2\theta_W \cos(2\beta)
\sin (\alpha+\beta) \right) \left(C_1(6)+C_{11}(6)+
C_{12}(6)\right)
\nonumber \\
&&
-\sin (\alpha-\beta) \sin (\alpha+\beta) \sin(2\beta)
\sec^2\theta_W C_1(4)
\nonumber \\
&&
+2 \cos(\alpha-\beta) \sin (\alpha-\beta) \left(C_2(3)-C_2(4) \right)
\nonumber \\
&&
-\cos (\alpha-\beta) \left(
2 \sin (\alpha-\beta)+\sec^2\theta_W \cos(\alpha+\beta)\sin(2\beta)
\right) \left( C_{11}(3)+C_{12}(3) \right)
\nonumber \\
&&
\left.
+\sin (\alpha-\beta) \left(
2 \cos (\alpha-\beta)-\sec^2\theta_W \sin(\alpha+\beta)\sin(2\beta)
\right) \left( C_{11}(4)+C_{12}(4) \right)
\right],
\end{eqnarray}
\begin{eqnarray}
f_{4}^{(2),\gamma,B}&=&
f_{5}^{(2),\gamma,B}
=f_{6}^{(2),\gamma,B}
=f_{3}^{(2),\gamma,B},
\end{eqnarray}
\begin{eqnarray}
f_{1}^{(2),Z,B}&=&
{
\alpha^2 m_W \over  2 \sin^2\theta_W (S-m_Z^2) }
\left[
{1 \over \cos^2\theta_W} 
\cos(\alpha-\beta) \sin(\alpha-\beta)
\left( -B_0(S,m_H^2,m_Z^2)+B_0(S,m_{h_0}^2,m_Z^2) \right)
\right.
\nonumber \\
&&
-{\cos(\alpha-\beta) 
\over  \cos^2\theta_W} 
\left(
\sin (\alpha-\beta) (m_Z^2+2 T+2 U)
+m_Z^2 \tan^2\theta_W\cos (\alpha+\beta)\sin(2\beta)
\right)C_0(13)
\nonumber \\
&&
+{ \sin(\alpha-\beta) 
\over \cos^2\theta_W} 
\left(
\cos (\alpha-\beta) (m_Z^2+2 T+2 U)
-m_Z^2 \tan^2\theta_W\sin (\alpha+\beta)\sin(2\beta)
\right)C_0(14)
\nonumber \\
&&
- \cos(\alpha-\beta) 
\left(
\sin (\alpha-\beta) (m_Z^2- S-m_H^2)
+m_Z^2 \tan^2\theta_W\cos (\alpha+\beta)\sin(2\beta)
\right)C_0(3)
\nonumber \\
&&
+\sin(\alpha-\beta) 
\left(
\cos (\alpha-\beta) (m_Z^2- S-m_{h_0}^2)
-m_Z^2 \tan^2\theta_W\sin (\alpha+\beta)\sin(2\beta)
\right)C_0(4)
\nonumber \\
&&
+  
(-2 m_{H^\pm}^2+T+U) \cos (\alpha-\beta)
\sin(\alpha-\beta) \left( C_1(3)-C_1(4) \right)
\nonumber \\
&&
+{1  
\over \cos^2\theta_W } 
(2 m_{H^\pm}^2+T+U) \cos (\alpha-\beta)
\sin(\alpha-\beta) \left( C_1(14)-C_1(13) \right)
\nonumber \\
&&
+  
(-2 m_{W}^2+T+U) \cos (\alpha-\beta)
\sin(\alpha-\beta) \left( C_2(4)-C_2(3) \right)
\nonumber \\
&&
+{1  
\over \cos^2\theta_W } 
(2 m_{W}^2+T+U) \cos (\alpha-\beta)
\sin(\alpha-\beta) \left( C_2(14)-C_2(13) \right)
\nonumber \\
&&
+{(\cos^2\theta_W-\sin^2\theta_W) 
\over \cos^2\theta_W } 
\sin(\alpha-\beta)
\left(-2 \cos (\alpha-\beta)
+\sec^2\theta_W\cos(\alpha+\beta)\cos(2\beta)
\right) C_{00}(5)
\nonumber \\
&&
-{(\cos^2\theta_W-\sin^2\theta_W) 
\cos(\alpha-\beta)
\over \cos^2\theta_W\sin^2\theta_W } 
\left(2  \sin (\alpha-\beta)
+\sec^2\theta_W\cos (\alpha+\beta) \sin(2\beta) \right)
C_{00}(3)
\nonumber \\
&&
+{  
\sin(\alpha-\beta)
\over \cos^4\theta_W } 
\left(-2 \sec^2\theta_W 
\cos (\alpha-\beta) +
\cos(2\beta) \cos(\alpha+\beta) \right)
C_{00}(7)
\nonumber \\
&&
-{  
\cos (\alpha-\beta)
\over \cos^4\theta_W } 
\left(-2 \sec^2\theta_W 
\sin (\alpha-\beta) +
\cos(2\beta) \sin (\alpha+\beta) \right)
C_{00}(8)
\nonumber \\
&&
-{ (\sin^2\theta_W-\cos^2\theta_W) 
\over \cos^4\theta_W } 
\cos (\alpha-\beta)
\sin (\alpha-\beta) 
\left( C_{00}(9)- C_{00}(10) \right)
\nonumber \\
&&
+{(\cos^2\theta_W-\sin^2\theta_W) 
\over \cos^2\theta_W } 
\cos(\alpha-\beta)
\left(-2 \sin (\alpha-\beta)
+\sec^2\theta_W\sin(\alpha+\beta)\cos(2\beta)
\right) C_{00}(6)
\nonumber \\
&&
-{ (\cos^2\theta_W-\sin^2\theta_W) 
\sin(\alpha-\beta)
\over \cos^4\theta_W } 
\left(2 \cos^2\theta_W \cos (\alpha-\beta)
-\sin (\alpha+\beta) \sin(2\beta) \right)
C_{00}(4)
\nonumber \\
&&
-{2  
\over \cos^2\theta_W } 
\cos(\alpha-\beta)
\sin(\alpha-\beta)
\left( C_{00}(11)- C_{00}(12) \right)
\nonumber \\
&&
+{  
\cos(\alpha-\beta)
\over \cos^2\theta_W } 
\left(  
2 \sin (\alpha-\beta) +
\sec^2\theta_W \sin(2\beta) \cos(\alpha+\beta) \right)
C_{00}(13)
\nonumber \\
&&
+{  
\sin (\alpha-\beta)
\over \cos^2\theta_W } 
\left(-2  \cos (\alpha-\beta) +
\sec^2\theta_W \sin (2\beta) \sin (\alpha+\beta) \right)
C_{00}(14),
\end{eqnarray}
\begin{eqnarray}
f_{2}^{(2),Z,B}&=&
{
-1+2 \sin^2\theta_W 
\over 
2 \sin^2\theta_W 
} f_{1}^{(2),Z,B},
\end{eqnarray}
\begin{eqnarray}
f_{3}^{(2),Z,B}&=&
{
\alpha^2 m_W \over  2 \sin^2\theta_W (S-m_Z^2) }
\left[
{2 \over \cos^2\theta_W} 
\cos(\alpha-\beta) \sin(\alpha-\beta)
\left(C_0(11)-C_0(12) +
\sec^2\theta_W 
\left( C_0(14)-C_0(13) \right)
\right) \right.
\nonumber \\
&&
+{ (\cos^2\theta_W-\sin^2\theta_W)
\over \cos^4\theta_W }
\sin(\alpha-\beta)
\left(
-2 \cos^2\theta_W \cos(\alpha-\beta)+
\cos(2\beta) \cos(\alpha+\beta)
\right)\times\nonumber 
\\
&& \left(C_1(5)+C_{11}(5)+C_{12}(5) \right)
+{ (\cos^2\theta_W-\sin^2\theta_W)
\over \cos^4\theta_W }
\cos (\alpha-\beta)
\cos(\alpha+\beta) \sin(2\beta)
C_1(3)
\nonumber \\
&&
+{1
\over \cos^4\theta_W }
\sin(\alpha-\beta)
\left(2 \cos^2\theta_W 
\cos(\alpha-\beta)-
 \cos (2\beta) \cos(\alpha+\beta)
\right)\left(C_1(7)+C_{11}(7)+C_{12}(7) \right)
\nonumber \\
&&
+{1
\over \cos^4\theta_W }
\cos(\alpha-\beta)
\left(-2 \cos^2\theta_W
\sin(\alpha-\beta)+
 \cos (2\beta) \sin(\alpha+\beta)
\right) \left(C_1(8)+C_{11}(8)+C_{12}(8) \right)
\nonumber \\
&&
+{ (\cos^2\theta_W-\sin^2\theta_W)
\over \cos^4\theta_W }
\cos(\alpha-\beta)
\sin(\alpha-\beta)
\left(C_1(9)-C_1(10)-C_{11}(9)+C_{11}(10)-C_{12}(9)
\right.\nonumber \\
&&\left.+C_{12}(10) \right)
+{\cos^2\theta_W-\sin^2\theta_W \over 
\cos^4\theta_W}
\cos (\alpha-\beta)
\left(
2 \cos^2 \theta_W \sin(\alpha-\beta)
- \cos(2\beta)  \sin(\alpha+\beta)
\right)\nonumber \\
&&
\left( C_1(6)+C_{11}(6)+C_{12}(6) \right)
+{ \cos^2\theta_W-\sin^2\theta_W
\over \cos^4\theta_W }
\sin(\alpha-\beta)
\sin(\alpha+\beta)
\sin(2\beta)
C_1(4)
\nonumber \\
&&
+{2 
\over \cos^2\theta_W }
\sin(\alpha-\beta)
\cos(\alpha-\beta)
\left(2 C_1(11)-2 C_1(12)+C_{11}(11)-C_{11}(11)
+C_{12}(11)-C_{12}(11) \right)
\nonumber \\
&&
-{1 
\over \cos^4\theta_W }
\cos(\alpha-\beta)
\left(
4 \cos^2\theta_W 
\sin(\alpha-\beta)+
\cos(\alpha+\beta)
\sin(2\beta)
\right)
C_1(13)
\nonumber \\
&&
-{1 
\over \cos^4\theta_W }
\sin(\alpha-\beta)
\left(
-4 \cos^2\theta_W 
\cos(\alpha-\beta)+
\sin(\alpha+\beta)
\sin(2\beta)
\right)
C_1(14)
\nonumber \\
&&
+4 
\cos(\alpha-\beta)
\sin(\alpha-\beta) 
\left(
C_2(4)-C_2(3)
\right)
\nonumber \\
&&
+{2  
\over \cos^2\theta_W}
\cos(\alpha-\beta)
\sin(\alpha-\beta) 
\left(
C_2(11)-C_2(12)+C_2(13)-C_2(14)
\right)
\nonumber \\
&&
+{ (\cos^2\theta_W-\sin^2\theta_W)
\over \cos^4\theta_W }
\cos (\alpha-\beta)
\left(
\cos(\alpha+\beta) \sin(2\beta)
+2 \cos^2\theta_W \sin(\alpha+\beta)
\right)
\left(C_{11}(3)+C_{12}(3) \right)
\nonumber \\
&&+{\cos^2\theta_W-\sin^2\theta_W
\over \cos^4\theta_W }
\sin(\alpha-\beta)
\left(
\sin(\alpha+\beta) \sin(2\beta)
-2 \cos^2\theta_W \sin(\alpha-\beta) \right)
\left(
C_{11}(4)+C_{12}(4)
\right)
\nonumber \\
&&
-{1 
\over \cos^4\theta_W }
\cos(\alpha-\beta)
\left(
2 \cos^2\theta_W 
\sin(\alpha-\beta)+
\cos(\alpha+\beta)
\sin(2\beta)
\right)
\left(
C_{11}(13)+C_{12}(13)
\right)
\nonumber \\
&&
\left.
-{1 
\over \cos^4\theta_W }
\sin(\alpha-\beta)
\left(
-2 \cos^2\theta_W 
\cos(\alpha-\beta)+
\sin(\alpha+\beta)
\sin(2\beta)
\right)
\left(
C_{11}(14)
+C_{12}(14)
\right)
\right],
\end{eqnarray}
\begin{eqnarray}
f_{4}^{(2),Z,B}&=&
f_{6}^{(2),Z,B}=
{
-1+2 \sin^2\theta_W 
\over 
2 \sin^2\theta_W 
} f_{3}^{(2),Z,B},
\end{eqnarray}
\begin{eqnarray}
f_{5}^{(2),Z,B}&=&
f_{3}^{(2),Z,B},
\end{eqnarray}
\begin{eqnarray}
f_1^{(3), \gamma}&=&{\alpha^2  
 \over 
4\sin^2\theta_W \cos^2\theta_W (m_{W}^2-m_{H^\pm}^2) m_W  S}
\Pi_{G^{\pm} H^\pm} \nonumber \\
&&-{\alpha^2  
(m_W^2-S)
 \over 
4\sin^2\theta_W \cos\theta_W (m_{W}^2-m_{H^\pm}^2) m_W m_{H^\pm}^2   S}
\Pi_{W^{\pm} H^\pm},
\end{eqnarray}
\begin{eqnarray}
f_1^{(3), Z}&=&{ \alpha^2 m_Z   
 \over 
4 \cos^3\theta_W (m_{W}^2-m_{H^\pm}^2) m_W^2  
(S-m_Z^2) }
\Pi_{G^{\pm} H^\pm} \nonumber \\
&&+
{\alpha^2  
(m_W^2-S) 
 \over 
4\sin^2\theta_W \cos\theta_W (m_{W}^2-m_{H^\pm}^2) m_W m_{H^\pm}^2   
(S- m_Z^2)}
\Pi_{W^{\pm} H^\pm},
\end{eqnarray}
\begin{eqnarray}
f_1^{(4),\gamma}&=&{\alpha^2 \left( m_W m_Z \sin(2 \beta) \sin(\alpha+\beta) 
-\cos\theta_W S \cos(\alpha-\beta) \right) \over 
8 \sin^2\theta_W \cos^2\theta_W m_{h_0}^2 m_W (m_W^2-m_{H^\pm}^2) S}
\Pi^T_{h_0} \nonumber \\
&&-{\alpha^2 \left( m_W m_Z \sin(2 \beta) \cos(\alpha+\beta) 
+\cos\theta_W S \sin(\alpha-\beta) \right) \over 
8\sin^2\theta_W \cos^2\theta_W m_{H}^2 m_W (m_W^2-m_{H^\pm}^2)  S}
\Pi^T_{H},
\end{eqnarray}
\begin{eqnarray}
f_1^{(4),Z}&=&{\alpha^2  \left( \cos(\alpha-\beta) \cos^2\theta_W (
m_Z^2 - S) - m_Z^2 \sin^2\theta_W \sin(2 \beta) \sin(\alpha+\beta) 
\right) \over 
8\sin^2\theta_W \cos^3\theta_W m_{h_0}^2 m_W (m_W^2-m_{H^\pm}^2)
( m_Z^2-S)}
\Pi^T_{h_0} \nonumber \\
&&+{\alpha^2  \left( \sin(\alpha-\beta) \cos^2\theta_W (
m_Z^2 - S) + m_Z^2 \sin^2\theta_W \sin(2 \beta) \cos(\alpha+\beta) 
\right) \over 
8 \sin^2\theta_W \cos^3\theta_W m_{H}^2 m_W (m_W^2-m_{H^\pm}^2)(
m_Z^2-S)}
\Pi^T_{H},
\end{eqnarray}
\begin{eqnarray}
f_1^{(5), \gamma}&=&-{\alpha^2  
\cos(\alpha-\beta)  \over 
8\sin^2\theta_W \cos\theta_W m_{h_0}^2 m_W S}
\Pi^T_{h_0} 
-{\alpha^2  
\sin(\alpha-\beta)  \over 
8\sin^2\theta_W \cos\theta_W m_{H}^2 m_W  S}
\Pi^T_{H},
\end{eqnarray}
\begin{eqnarray}
f_1^{(5),Z}&=&{\alpha^2 \sin^2\theta_W  \cos(\alpha-\beta) 
\over 
8 \cos^3\theta_W m_{h_0}^2 m_W (
+m_Z^2-S)}
\Pi^T_{h_0} 
+{\alpha^2 \sin^2\theta_W  \sin(\alpha-\beta) 
\over 
8 \cos^3\theta_W m_{H}^2 m_W  (
m_Z^2-S)}
\Pi^T_{H},
\end{eqnarray}
\begin{eqnarray}
f_{1}^{(6),\gamma}&=&
{\alpha^2 m_Z \over 4 \sin^2\theta_W\cos\theta_W S}
\left[
\sin(\alpha-\beta)
\left(
2 \sec^2\theta_W \cos(\alpha-\beta)
- \cos(2\beta) \cos(\alpha+\beta)
\right) B_0(m_{H^\pm}^2,m_H^2,m_{H^\pm}^2)
\right.
\nonumber \\
&&
-\cos(\alpha-\beta)
\left(
\sec^2\theta_W \sin(\alpha-\beta)
+\sin(2\beta) \cos(\alpha+\beta)
\right) B_0(m_{H^\pm}^2,m_H^2,m_{W}^2)
\nonumber \\
&&
+\cos(\alpha-\beta)
\left(
-2 \sec^2\theta_W \sin(\alpha-\beta)
+\cos(2\beta) \sin(\alpha+\beta)
\right) B_0(m_{H^\pm}^2,m_{H^\pm}^2,m_{h_0}^2)
\nonumber \\
&&
\left.
+\sin(\alpha-\beta)
\left(
\sec^2\theta_W \cos(\alpha-\beta)
- \sin(2\beta) \sin(\alpha+\beta)
\right) B_0(m_{H^\pm}^2,m_{h_0}^2,m_{W}^2)
\right],
\end{eqnarray}
\begin{eqnarray}
f_{1}^{(6),Z}&=&
{\alpha^2 m_Z \over 4 \cos^3\theta_W (S-m_Z^2)}
\left[
\sin(\alpha-\beta)
\left(
2\sec^2\theta_W \cos(\alpha-\beta)
- \cos(2\beta) \cos(\alpha+\beta)
\right) B_0(m_{H^\pm}^2,m_H^2,m_{H^\pm}^2)
\right.
\nonumber \\
&&
-\cos(\alpha-\beta)
\left(
\sec^2\theta_W \sin(\alpha-\beta)
+\sin(2\beta) \cos(\alpha+\beta)
\right) B_0(m_{H^\pm}^2,m_H^2,m_{W}^2)
\nonumber \\
&&
+\cos(\alpha-\beta)
\left(
-2 \sec^2\theta_W \sin(\alpha-\beta)
+\cos(2\beta) \sin(\alpha+\beta)
\right) B_0(m_{H^\pm}^2,m_{H^\pm}^2,m_{h_0}^2)
\nonumber \\
&&
\left.
+\sin(\alpha-\beta)
\left(
\sec^2\theta_W \cos(\alpha-\beta)
- \sin(2\beta) \sin(\alpha+\beta)
\right) B_0(m_{H^\pm}^2,m_{h_0}^2,m_{W}^2)
\right],
\end{eqnarray}
\begin{eqnarray}
f_{1}^{(7),\gamma}&=&
{ \alpha^2 m_W \over 2 \sin^2\theta_W S}
\cos(\alpha-\beta)
\sin(\alpha-\beta)
\left(
B_0(m_W^2,m_H^2,m_W^2)
-B_0(m_W^2,m_{h_0}^2,m_W^2)
\right),
\end{eqnarray}
\begin{eqnarray}
f_{1}^{(7),Z}&=&
{ \alpha^2 m_W \over 2 \cos^2\theta_W (S-m_Z^2)}
\cos(\alpha-\beta)
\sin(\alpha-\beta)
\left(
B_0(m_W^2,m_H^2,m_W^2)
-B_0(m_W^2,m_{h_0}^2,m_W^2)
\right),
\end{eqnarray}
\begin{eqnarray}
f_{1}^{(8),Z}&=&
{\alpha^2 m_Z \over 2 \cos^3\theta_W (S-m_Z^2)}
\cos(\alpha-\beta)
\sin(\alpha-\beta)
\left(
B_0(S,m_H^2,m_Z^2) 
-B_0(S,m_{h_0}^2,m_Z^2)
\right),
\end{eqnarray}
\begin{eqnarray}
f_{1}^{(9),\gamma}&=&
{ \alpha^2 m_W \over 8 \cos^2\theta_W \sin^2\theta_W
(m_{H^\pm}^2-m_W^2) S}
\left[
\cos(2\beta) \sin(2\beta)
\left( 
A_0(m_{A_0}^2)
+4 A_0(m_{H^\pm}^2)
-A_0(m_{Z}^2)
-4 A_0(m_{W}^2) \right)
\right.
\nonumber \\
&&
\left.
-\left(
\cos^2\theta_W \cos(2\beta) \sin(2\alpha)
+\sin^2\theta_W \cos(2\alpha) \sin(2\beta)
\right)
\left(
A_0(m_H^2)
-A_0(m_{h_0}^2)
\right)
\right],
\end{eqnarray}
\begin{eqnarray}
f_{1}^{(9),Z}&=&
{ \alpha^2 m_W \over 8 \cos^4\theta_W 
(m_{H^\pm}^2-m_W^2) (S-m_Z^2)}
\left[
\cos(2\beta) \sin(2\beta)
\left( 
A_0(m_{A_0}^2)
+4 A_0(m_{H^\pm}^2)
-A_0(m_{Z}^2)
-4 A_0(m_{W}^2) \right)
\right.
\nonumber \\
&&
\left.
-\left(
\cos^2\theta_W \cos(2\beta) \sin(2\alpha)
+\sin^2\theta_W \cos(2\alpha) \sin(2\beta)
\right)
\left(
A_0(m_H^2)
-A_0(m_{h_0}^2)
\right)
\right],
\end{eqnarray}
\begin{eqnarray}
f_2^{(j),\gamma}= f_1^{(j),\gamma}, \mbox{ for} \ j=3-9,
\end{eqnarray}
\begin{eqnarray}
f_2^{(j),Z}= {2 \sin^2\theta_W
\over -1+2 \sin^2\theta_W}
f_1^{(j),Z}, \mbox{ for} \ j=3-9,
\end{eqnarray}
\begin{eqnarray}
f_2^{(10)}=-{ \alpha^2  
 \over 
8 \sin\theta_W \cos\theta_W (m_{W}^2-m_{H^\pm}^2) m_W m_{H^\pm}^2  \pi^2}
\Pi_{W^{\pm} H^\pm},
\end{eqnarray}
\begin{eqnarray}
f_2^{(11)}&=&{\alpha^2  
\cos(\alpha-\beta)  \over 
16 \sin^2\theta_W \cos\theta_W m_{h_0}^2 m_W (m_W^2-m_{H^\pm}^2)}
\Pi^T_{h_0} \nonumber \\
&&+{\alpha^2  
\sin(\alpha-\beta)  \over 
16\sin^3\theta_W \cos\theta_W m_{H}^2 m_W (m_W^2-m_{H^\pm}^2) }
\Pi^T_{H},
\end{eqnarray}
where
\begin{eqnarray}
\Pi^T_{h_0} &=& 
-24 \cos \theta_W m_t^2 \cos\alpha \csc\beta A_0(m_t^2)
+24 \cos \theta_W m_b^2 \sin\alpha \sec\beta A_0(m_b^2)
\nonumber \\
&&
-m_W m_Z \left(6 \sin(\alpha-\beta)+ \sin(\alpha+\beta)\cos(2 \beta)
\right)
A_0(m_Z^2)
\nonumber \\
&&
-2 m_W \left( 6 \cos\theta_W m_W \sin(\alpha-\beta)
+m_Z \cos(2 \beta) \sin (\alpha+\beta) \right) A_0(m_W^2)
\nonumber \\
&&
-m_W m_Z \left(\cos(2 \alpha) \sin(\alpha+\beta)+2 \cos(\alpha+\beta)
\sin(2 \alpha)\right) A_0(m_H^2)
\nonumber \\
&&
+3 m_W m_Z \cos(2 \alpha) \sin(\alpha+\beta) A_0(m_{h_0}^2)
+m_W m_Z \cos(2\beta) \sin(\alpha+\beta) A_0(m_{A_0}^2)
\nonumber \\
&&
+2 m_W \left(m_Z \cos(2 \beta) \sin(\alpha+\beta)
-2 \cos\theta_W m_W \sin(\alpha-\beta)\right) A_0(m_{H^\pm}^2),
\end{eqnarray}
\begin{eqnarray}
\Pi^T_{H} &=& 
-24 \cos \theta_W m_t^2 \sin\alpha \csc\beta A_0(m_t^2)
-24 \cos \theta_W m_b^2 \cos\alpha \sec\beta A_0(m_b^2)
\nonumber \\
&&
+m_W m_Z \left(6 \cos(\alpha-\beta)+ \cos(\alpha+\beta)\cos(2 \beta)
\right)
A_0(m_Z^2)
\nonumber \\
&&
+2 m_W \left( 6 \cos\theta_W m_W \cos(\alpha-\beta)
+m_Z \cos(2 \beta) \cos (\alpha+\beta) \right) A_0(m_W^2)
\nonumber \\
&&
+3 m_W m_Z \left(\cos(2 \alpha) \cos(\alpha+\beta)
 \right) A_0(m_H^2)
-m_W m_Z \cos(2\beta) \cos(\alpha+\beta) A_0(m_{A_0}^2)
\nonumber \\
&&
+ m_W m_Z \left( 2 \sin(2 \alpha) \sin(\alpha+\beta)
-\cos(2 \alpha) \cos(\alpha+\beta) \right) A_0(m_{h_0}^2)
\nonumber \\
&&
-2 m_W \left(m_Z \cos(2 \beta) \cos(\alpha+\beta)
-2 \cos\theta_W m_W \cos(\alpha-\beta)\right) A_0(m_{H^\pm}^2),
\end{eqnarray}
\begin{eqnarray}
\Pi_{G^{\pm} H^{\pm}} &=& 
12 \cos^2\theta_W (m_t^2 \cot\beta-m_b^2 \tan\beta) A_0(m_t^2)
\nonumber \\
&&
+ 6 \cos^2\theta_W (m_t^2 -m_b^2) (m_b^2 \tan\beta- m_t^2 \cot\beta)
B_0(0,m_b^2, m_t^2)
\nonumber \\
&&
+ 6 \cos^2\theta_W \left(
\cot\beta m_t^2 (m_t^2-m_b^2-m_{H^\pm}^2)
+\tan\beta m_b^2 (m_t^2-m_b^2+m_{H^\pm}^2)
\right) B_0(m_{H^\pm}^2,m_b^2, m_t^2)
\nonumber \\
&&
+ \cos^2\theta_W (m_{h_0}^2-m_W^2) m_W^2 \cos(\alpha-\beta)
\sin(\alpha-\beta) B_0(0,m_{h_0}^2,m_W^2)
\nonumber \\
&&
-\cos^2\theta_W (m_{H}^2-m_W^2) m_W^2 \cos(\alpha-\beta)
\sin(\alpha-\beta) B_0(0,m_{H}^2,m_W^2)
\nonumber \\
&&
+m_W^2 \left[2 \cos\theta_W m_W \cos(\alpha-\beta) - 
m_Z \cos (2 \beta) \cos (\alpha+\beta)\right]
\left[ \cos\theta_W m_W \sin(\alpha-\beta)
\right.
\nonumber \\
&& 
\left. +m_Z \sin(2 \beta) \cos (\alpha+\beta)\right]
B_0(m_{H^\pm}^2,m_H^2,m_{H^\pm}^2)
\nonumber \\
&&
+m_W^2 \left[ m_Z^2 \cos(2\beta) \cos^2(\alpha+\beta)
\sin(2 \beta) +\cos^2\theta_W \sin(\alpha-\beta) \left(\right.
\right.
\nonumber \\
&&\left.\left. \cos(\alpha-\beta) (2 m_H^2+2 m_{H^\pm}^2
-m_W^2)+m_Z^2 \cos(2 \beta) \cos(\alpha+\beta) \right) \right]
B_0(m_{H^\pm}^2,m_H^2,m_{W}^2)
\nonumber \\
&&
+m_W^2 \left[-2 \cos\theta_W m_W \sin(\alpha-\beta) + 
m_Z \cos (2 \beta) \sin (\alpha+\beta)\right]
\left[ \cos\theta_W m_W \cos (\alpha-\beta)
\right.
\nonumber \\
&& 
\left. -m_Z \sin(2 \beta) \sin (\alpha+\beta)\right]
B_0(m_{H^\pm}^2,m_{H^\pm}^2,m_{h_0}^2)
\nonumber \\
&&
+m_W^2 \left[ m_Z^2 \sin(2 \beta) \cos(2\beta) \sin^2(\alpha+\beta)
+\cos^2\theta_W \cos(\alpha-\beta) \left(\right.
\right.
\nonumber \\
&&\left.\left. \sin(\alpha-\beta) (-2 m_{h_0}^2-2 m_{H^\pm}^2
+m_W^2)-m_Z^2 \sin(\alpha+\beta)\cos(2 \beta) \right) \right]
B_0(m_{H^\pm}^2,m_{h_0}^2,m_{W}^2),
\end{eqnarray}
\begin{eqnarray}
\Pi_{W^{\pm} H^{\pm}} &=& 
 6 \cos\theta_W (m_b^2 -m_t^2) (m_b^2 \tan\beta+ m_t^2 \cot\beta)
B_0(0,m_b^2, m_t^2) \nonumber \\
&&
+6 \cos\theta_W \left(
\cot\beta m_t^2 (m_t^2-m_b^2-m_{H^\pm}^2) +
\tan\beta (m_b^2 (m_t^2-m_b^2+m_{H^\pm}^2) \right)
B_0(m_{H^\pm}^2,m_b^2, m_t^2) \nonumber \\
&&
+\sin(\alpha-\beta)m_W (m_{H^\pm}^2-m_H^2)
\left(m_Z \cos(2 \beta) \cos(\alpha+\beta)-
2 \cos\theta_W m_W \cos (\alpha-\beta)\right)\times
\nonumber \\
&&
\left(B_0(0,m_H^2,m_{H^\pm}^2)-B_0(m_{H^\pm}^2,m_H^2,m_{H^\pm}^2) \right)
+\cos(\alpha-\beta) m_W (m_{W}^2-m_H^2)\times
\nonumber \\
&&
\left(m_Z \sin(2 \beta) \cos(\alpha+\beta)+
2 \cos\theta_W m_W \sin (\alpha-\beta)\right)
B_0(0,m_H^2,m_{W}^2)
\nonumber \\
&&+
m_W \cos(\alpha-\beta) \left[ \sin (\alpha-\beta)
\cos\theta_W m_W (2 m_H^2+3 m_{H^\pm}^2-2 m_W^2)
\right.
\nonumber \\
&&+\left.
\cos(\alpha+\beta) \sin(2 \beta)
m_Z (m_H^2-m_W^2) \right]
B_0(m_{H^\pm}^2,m_H^2,m_{W}^2)
\nonumber \\
&&
-\cos(\alpha-\beta)m_W (m_{H^\pm}^2-m_{h_0}^2)
\left(m_Z \cos(2 \beta) \sin(\alpha+\beta)-
2 \cos\theta_W m_W \sin (\alpha-\beta)\right)\times
\nonumber \\
&&
\left(B_0(0,m_{h_0}^2,m_{H^\pm}^2)-B_0(m_{H^\pm}^2,m_{h_0}^2,m_{H^\pm}^2) 
\right)
+\sin(\alpha-\beta) m_W (m_{W}^2-m_{h_0}^2)\times
\nonumber \\
&&
\left(m_Z \sin(2 \beta) \sin(\alpha+\beta)-
2 \cos\theta_W m_W \cos (\alpha-\beta)\right)
B_0(0,m_{h_0}^2,m_{W}^2)
\nonumber \\
&&+
m_W \sin(\alpha-\beta) \left[ \cos (\alpha-\beta)
\cos\theta_W m_W (-2 m_{h_0}^2-3 m_{H^\pm}^2+2 m_W^2)
\right.
\nonumber \\
&&+\left.
\sin(\alpha+\beta) \sin(2 \beta)
m_Z (m_{h_0}^2-m_W^2) \right]
B_0(m_{H^\pm}^2,m_{h_0}^2,m_{W}^2).
\end{eqnarray}
In above equations, the definitions of 
$B_0$, $C$ and $D$ functions 
could be found in Ref. \cite{denner}, 
 the variables
of $C$ functions from 1 to 14 represent
$
(m_{H^\pm}^2,S,m_W^2,i)$, where $i$ stand for the internal 
mass square of the loops, for $i=1-14$, which are
\begin{eqnarray}
&&(m_t^2,m_b^2,m_b^2), (m_b^2,m_t^2,m_t^2), 
(m_H^2,m_W^2,m_W^2), (m_{h_0}^2,m_W^2,m_W^2)
\nonumber \\
&&(m_H^2,m_{H^\pm}^2,m_{H^\pm}^2),
(m_{h_0}^2,m_{H^\pm}^2,m_{H^\pm}^2),
(m_{H^\pm}^2,m_H^2,m_{A_0}^2),
(m_{H^\pm}^2,m_{h_0}^2,m_{A_0}^2),
\nonumber \\
&&(m_{H^\pm}^2,m_Z^2,m_H^2),
(m_{H^\pm}^2,m_Z^2,m_{h_0}^2),
(m_{W}^2,m_{A_0}^2,m_H^2),
(m_{W}^2,m_{A_0}^2,m_{h_0}^2),
\nonumber \\
&&(m_W^2,m_H^2,m_Z^2),
(m_W^2,m_{h_0}^2,m_Z^2),
\end{eqnarray}
 and the variables
of $D$ functions from 1 to 2 represent
$
(m_{H^\pm}^2,0,0,m_W^2,T,S,i)$, where $i$ stand for the internal 
mass square of the loops, for $i=1-2$, which are
\begin{eqnarray}
&&(m_H^2,m_W^2,0,m_W^2), (m_{h_0}^2,m_W^2,0,m_W^2).
\end{eqnarray}
 
\end{document}